# Metallic temperature dependence of resistivity in perchlorate doped polyacetylene

Y. W. Park, E. S. Choi, D. S. Suh

Department of Physics and Condensed Matter Research Institute, Seoul National University, Seoul 151-742, Korea

**Abstract**

We have measured the electrical resistivity ($\rho$) and the thermoelectric power (TEP) of the perchlorate ($ClO_4^-$) doped stretch oriented polyacetylene (PA) film. For the highly conducting samples ($\sigma_{RT} \approx 41000$ S/cm), the temperature dependence of the 4-probe resistivity shows positive temperature coefficient of resistivity (TCR) from T=1.5K to 300K. For the less conducting samples, the 4-probe resistivity data show the crossover of TCR with a broad minimum peak at T=T*$\approx$200K. For samples of $\sigma_{RT} \geq 20000$ S/cm, the $\rho(1.5K)/\rho(300K)<1$, i.e., the resistivity at 1.5K is lower than the room temperature resistivity value. The temperature dependence of the TEP shows diffusive linear metallic TEP becoming temperature independent below 40K. Unlike the others who used $Cu(ClO_4)_2$ for the $ClO_4^-$ doping, the initial doping material we used is anhydrous $Fe(ClO_4)_3$ which is crucial to obtain the positive TCR from T=1.5K to 300K.

**Introduction**

From the first discovery of insulator-metal transition by doping [1], metallic polyacetylene (PA) has been widely studied theoretically and experimentally for many years. Metallic properties of heavily doped PA, such as temperature independent magnetic susceptibility, high electrical conductivity up $10^5$ S/cm at room temperature, and linear temperature dependent TEP [2-4], were observed. From the quasi 1-dimensional metallic nature of PA, the room temperature conductivity ($\sigma_{RT}$) was estimated to be greater than that of copper [5]. But even for the most highly conducting samples, the temperature dependence of resistivity is normally non-metallic; i.e., the temperature coefficient of resistivity (TCR) is negative. There have been some reports about the crossover of TCR from positive to negative near 200K for $AsF_5$, $FeCl_3$, $I_2$ and $ClO_4$ doped PA [6-10] and for the other conducting polymers at low temperature [11-13].

It is thought that the non-metallic temperature dependence is due to the substantial disorder which causes localization of the electron wave functions. From the synthesis and the sample preparation approaches, many efforts have been made to reduce the disorder in the sample to investigate the intrinsic metallic state of conducting polymers.

We have doped the stretch oriented PA film with $ClO_4^-$ and measured the electrical conductivity and the TEP. For samples of $\sigma_{RT}>20000$ S/cm, we observed $\rho(1.5K)/\rho(300K)<1$ and the crossover of TCR below 200K with much broader minimum peak than that of the previous reports [6-10]. In addition, for the first time, we discovered the positive TCR from T=1.5K to 300K for the most highly conducting $ClO_4^-$ doped PA ($\sigma_{RT} \approx 41000$ S/cm).

**Experimental**

The Naarmann type PA film was synthesized by the modified Shirakawa method with the catalyst condition of [Al]/[Ti]=2.0 [14]. The perchlorate doping was done by immersing the stretched film ($l/l_0=4$) in the 0.1M anhydrous $Fe(ClO_4)_3$ / acetonitrile solution. Anhydrous $Fe(ClO_4)_3$ was prepared by drying $Fe(ClO_4)_3 \cdot H_2O$ powder under dynamic vacuum while heating the powder in the doping apparatus with a heat gun. The doping concentration was controlled by adjusting the immersing time and the concentration of solution. After the film was doped to the saturation level, the doping concentration was determined by measuring the weight uptake of the sample. The ionic state of the dopant in the PA film analyzed by the Energy Dispersive Spectroscopy (EDS) as well as the ICP-Emission is identified as in the form of $(ClO_4)^-$ with Fe content less than 0.01 %. The Extended X-ray Absorption Fine Structure (EXAFS) results measured by synchrotron light source, shows a small hump at Fe K-edge energy indicating a small amount of (less than 0.01%) Fe trace in the doped film. The doping concentration is about 7~8 % for the samples doped to the saturation level. Anhydrous $Cu(ClO_4)_2$ was also used as an initial doping material. The doping process was similar to the $Fe(ClO_4)_3$ case and it is analyzed that the doped PA contains Cu less than 0.02 %. The $FeCl_3$ doping was